\documentclass[aps,prb,twocolumn,superscriptaddress,showpacs]{revtex4}
\usepackage{graphicx}
\usepackage{latexsym}
\usepackage{amsmath}
\usepackage{txfonts}
\usepackage{helvet}
\usepackage{braket}
\usepackage{amscd}

\begin{document}

\title{Orbital reformation with vanadium trimerization in $d^2$ triangular lattice LiVO$_2$ revealed by $^{51}$V NMR} 

\author{Takaaki Jin-no}
\affiliation{Department of Physics, Graduate School of Science, Nagoya University, Furo-cho, Chikusa-ku, Nagoya 464-8602, Japan}
\author{Yasuhiro Shimizu}
\affiliation{Department of Physics, Graduate School of Science, Nagoya University, Furo-cho, Chikusa-ku, Nagoya 464-8602, Japan}
\author{Masayuki Itoh}
\affiliation{Department of Physics, Graduate School of Science, Nagoya University, Furo-cho, Chikusa-ku, Nagoya 464-8602, Japan}
\author{Seiji Niitaka}
\affiliation{RIKEN Advanced Science Institute, 2-1, Hirosawa, Wako, Saitama 351-0198, Japan}
\author{Hidenori Takagi}
\affiliation{RIKEN Advanced Science Institute, 2-1, Hirosawa, Wako, Saitama 351-0198, Japan}
\date{\today}

\begin{abstract}
	LiVO$_2$ is a model system of the valence bond solid (VBS) in $3d^2$ triangular lattice. 
	The origin of the VBS formation has remained controversial. 
	We investigate the microscopic mechanism by elucidating the $d$ orbital character via on-site $^{51}$V NMR measurements in a single crystal up to 550 K across a structural transition temperature $T_c$. 
	The Knight shift, $K$, and nuclear quadrupole frequency, $\delta \nu$, show that the $3d$ orbital with local trigonal symmetry are reconstructed into a $d_{yz}d_{zx}$ orbital order below $T_c$. 
	Together with the NMR spectra with three-fold rotational symmetry, we confirm a vanadium trimerization with $d$-$d$ $\sigma$ bonds. 
	The Knight shift extracts the large Van-Vleck orbital susceptibility, $\chi^{\rm VV} = 3.6 \times 10^{-4}$, in a paramagnetic state above $T_c$, which is comparable to the spin susceptibility. 
	The results suggest that orbitally induced Peierls transition in the proximity of the frustrated itinerant state is the dominant driving force of the trimerization transition. 
	
\end{abstract}



\maketitle

	Valence bond ordering in two- and three-dimensional systems is a manifestation of geometrical frustration.\cite{Mendels} 
	Across Mott transition, a valence bond solid (VBS) phase resides nearby unconventional metal\cite{Shimizu3, Katayama} and superconducting phases\cite{Shimizu3} in triangular lattice systems. 
	The VBS transition can accompany by $d$ orbital orders with versatile textures, depending on lattice geometry and band filling.\cite{Goodenough,Pen,Radaelli,Schmidt,Khomskii,text2} 
	However, $d$ orbital orders governed by frustrated intersite interactions are not apparent from lattice distortions and need to be detected by microscopic probes. 
	Here we show a tool for detailed characterization of $d^2$ orbital ordering in a model material LiVO$_2$ with the triangular lattice. 

	$ABX_2$ ($A$:alkali metal; $B$: transition metal; $X$: chalcogen ion) with an ordered lock-type structure is a prototype of triangular-lattice antiferromagnet.\cite{text2} 
	For Ti$^{3+}$ and V$^{3+}$, $t_{2g}$ orbital degrees of freedom play crucial roles in determining magnetic ground states. 
	Among $ABX_2$, LiVO$_2$ has been extensively investigated as a candidate of VBS.\cite{Picciotto, Cardoso, Takei, Imai, Tian} 
	The crystal structure is comprised of the triangular lattice of V$^{3+}$ ($3d^2$) ions in a rhombohedral $R\bar{3}c$ structure [Fig. 1(a)]. 
	A trigonal elongation of VO$_6$ along the $c$ axis splits the $t_{2g}$ sublevel into a lower $a_{1g}$ singlet and an upper $e_g^\prime$ doublet [Figs. 1(b,c)].\cite{Ezhov} 
	Calculated orbital occupations are almost equivalent in the $a_{1g}$ and $e_g^\prime$ orbitals due to the small trigonal field of 25 meV compared to the bandwidth.\cite{Ezhov} 
	Below a first-order structural transition temperature $T_c \sim$ 450-500 K, a superlattice ${\surd 3}a \times {\surd 3}a \times 2c$ [Fig. 1(a)] appears in x-ray and electron diffraction measurements.\cite{Cardoso, Takei, Imai, Tian} 
	Recent pair distribution function analyses of synchrotron data propose lower-symmetry lattice distortions depending on the layers.\cite{Pourpoint} 
	Although the way of displacements of vanadium ions is compatible to trimerization,\cite{Takei, Imai, Pourpoint} the orbital character has not been accessible by the diffraction measurements due to Li deficiency and crystal twinning. 

	The magnetic susceptibility of LiVO$_2$ drops at $T_c$ with spin gap of several thousands Kelvin estimated from the $^{51}$V nuclear spin-lattice relaxation rate $T_1^{-1}$.\cite{Onoda, Tanaka}
	Hence the ground state is most likely nonmagnetic without residual paramagnetic spins. 
	The spin-singlet state in the $d^2$ triangular lattice has been theoretically explained as a spin-Peierls transition accompanied by a $d_{yz}d_{zx}$ orbital order in a Heisenberg model.\cite{Pen} 
	However, the conducting in-plane resistivity, $\sim 0.1 \Omega$cm, and the constant magnetic susceptibility\cite{Takei, Tian} above $T_c$ may be incompatible to a strongly localized picture. 
	Since the ground states are chemically tuned from a metallic LiVSe$_2$ (Ref. \onlinecite{Katayama}) to an antiferromagnetic insulator NaVO$_2$ (Ref. \onlinecite{Mcqueen}), a spin-singlet phase appear to be located in an intermediate parameter region as a function of electron correlation. 
	A Hubbard model calculation explains such behavior by involving a strong trigonal crystal field and also modifies the ground state for the Heisenberg limit to be a ferrimagnetic order.\cite{Yoshitake} 
	Thus, the origin of the spin-singlet state in LiVO$_2$ remains controversial and to be elucidated via direct observations of the orbital structure. 

	\begin{figure}
	\includegraphics[scale=0.6]{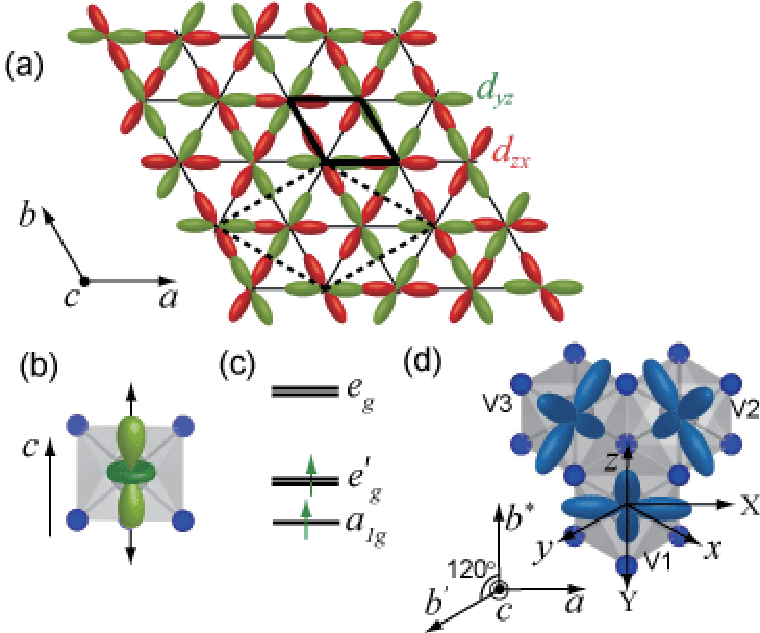}
	\caption{\label{Fig1} 
	(Color online) (a) Layered triangular lattice of LiVO$_2$ viewed from the $c$ axis with an orbital texture obtained from the present study, where only two of four leaves in $d_{yz}$ and $d_{zx}$ orbitals are shown for clarity. 
	Thick solid and dotted lines represent unit cells of an original $R\bar{3}m$ lattice above $T_c$ and a ${\surd 3}a \times {\surd 3}a$ superlattice below $T_c$, respectively (Refs.\onlinecite{Takei, Tian}).  
	(b) VO$_6$ unit trigonally elongated along $c$ axis above $T_c$. 
	(c) $d$ levels in a trigonal field: lowest $a_{1g}$ singlet, middle $e_g^\prime$ doublet, and higher $e_g$ doublet. 
	(d) Minority $d_{xy}$ orbitals that dominates the anisotropy of NMR spectra below $T_c$. 
	The $x$, $y$, and $z$ directions are taken parallel to the VO bond directions and identical to $Y-X$, $X+Y$, and $Z$ axes, where $X$, $Y$, and $Z$ are the principal axes of the local symmetry that determines the $K$ and $\delta \nu$ tensor. 
	The $b^*$ axis is defined perpendicular to the $a$ and $c$ axes. 
	The $b^\prime$ axis is located at 30$^\circ$ from the $a$ axis. 
	}
	\end{figure}

	In this work, we report detailed $^{51}$V NMR studies on a single crystal of LiVO$_2$, which provide microscopic insights into the spin-singlet transition. 
	Beyond previous $^{51}$V and $^7$Li NMR studies for powder samples, which lack information of $3d$ orbitals,\cite{Kikuchi, Pourpoint} we determine $d$ orbital characters by investigating anisotropic tensors of Knight shift, $K$, and nuclear quadrupole splitting frequency, $\delta \nu$. 
	The tensors satisfy an orbital order with the three-fold symmetry in the spin-singlet state, which promotes $\sigma$-bonding vanadium trimerization. 
	We also decompose the orbital and spin Knight shifts, which provides the significant residual orbital degeneracy. 

	\begin{figure*}
	\includegraphics[scale=0.68]{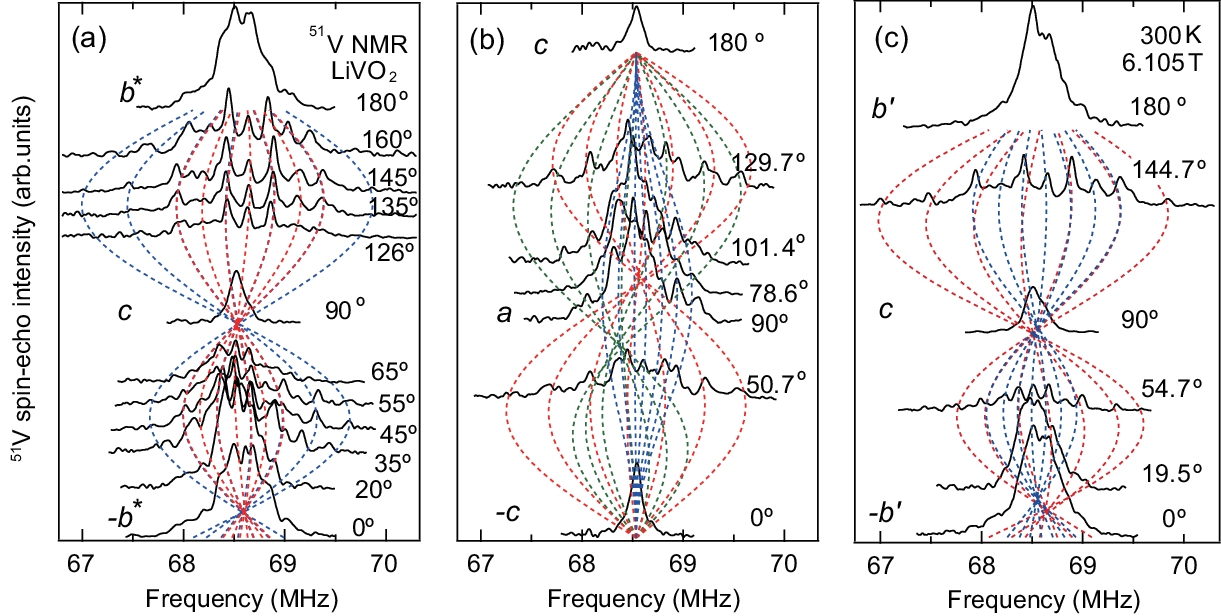}
	\caption{\label{Fig2}
	(Color online) Frequency-swept $^{51}$V NMR spectra in LiVO$_2$ (300 K, $B_0$ = 6.105 T). 
	Magnetic field is rotated around the (a) $a$, (b) $b^*$, and (c) $b$ axes. 
	Dotted guide lines are sinusoidal fitting\cite{Volkov} at the bottom of the peak position by taking the vertical axis as a rotation angle. 
	Blue, red, and green lines are assigned to V1, V2, and V3 in Fig. 1(c), respectively. 
	}
	\end{figure*}

	Single crystals of LiVO$_2$ were obtained from a LiBO$_2$-Li$_2$O flux method in a sealed quartz tube by using LiVO$_2$ powders prepared by a solid state reaction of Li$_2$CO$_3$ and V$_2$O$_3$.\cite{Tian} 
	$^{51}$V NMR spectra were obtained from the spin-echo measurements in a constant magnetic field $B_0$ = 6.105 T and a temperature range 290-550 K. 
	The angular dependences of $\delta \nu$ and $K$ were obtained for the single crystal in rotation planes including the principal axes by utilizing a two-axis goniometer at 300 K. 
	High-temperature experiments were performed for aligned single crystals with the uniaxial rotation in an oven, where the temperature was monitored by a thermocouple near the sample. 

	NMR is useful as a local probe of $d$ and $p$ orbital occupations in transition metal\cite{Abragam, Hanzawa, Zheng, Kiyama} and organic compounds\cite{Autschbach} via anisotropic electric and magnetic hyperfine interactions. 
	For $t_{2g}$ orbitals weakly bonded to ligand $p$ orbitals, the symmetry axis of $K$ and $\delta \nu$ reflects occupied $3d$ orbitals. 
	The method can be applied in the spin-singlet state.\cite{Shimizu} 
	For $^{51}$V nuclear spins, $I = 7/2$, electric quadrupole interactions under the electric field gradient (EFG) split NMR spectra with an interval frequency $\delta \nu$. 
	$\delta \nu$ consists of the on-site $d$ orbital and the outer ligand terms: $\delta \nu = \delta \nu^{\rm on} + \delta \nu^{\rm out}$, where the components of $\delta \nu^{\rm on}$ are expressed by\cite{Abragam} 
	\begin{eqnarray}
	\delta \nu^{\rm on}_{ij} = \frac{3e^2Q}{2I(2I-1)}\xi \braket {r^{-3}}q_{ij}, 
	\end{eqnarray} 
	where $e$ is the electron charge, $Q$ the nuclear quadrupole moment, $-0.05 \times 10^{-24}$ cm$^2$ for $^{51}$V, and $\braket {r^{-3}}$ the V$^{3+}$ radial expectation value with a reduction factor of 0.8 compared to the free ion value, $\xi=1/21$, and $q_{ij} \equiv \frac{1}{2}(L_i L_j + L_j L_i ) - \frac{1}{3}L(L+1)\delta_{ij}$ ($L$: orbital angular momentum). 
	$\delta \nu^{\rm on}_{ij}$ is proportional to the orbital polarization reflected in $q_{ij}$. 
	For instance, a full polarization of $d_{xy}$ gives the principal components $q_{xx}=q_{yy}=-1$ and $q_{zz}=2$.\cite{Abragam,Autschbach} 
	On the other hand, $\delta \nu^{\rm out}$ is given as\cite{Abragam}
	\begin{eqnarray}
	\delta \nu^{\rm out}_{ij} = \frac{(1 - \gamma_\infty )eQ}{6I(2I-1)}\sum_{ij} V_{ij} \Biggl[\frac{3}{2}(I_i I_j + I_j I_i) - \delta_{ij} I^2\Biggr], 
	\end{eqnarray} 
	where $\gamma_\infty$ is the Sternheimer antishielding factor due to core-shell polarization, and $V_{ij}$ the EFG tensor of surrounding ions.\cite{Abragam} 
	Magnetic hyperfine interactions between $^{51}$V nuclear spin and electrons produce Knight shift, where the components are derived from a spin-Hamiltonian expressed in a second-order approximation as\cite{Abragam} 
	\begin{equation}
	K_{ij} = \mathcal {P} (\kappa \delta_{ij} + 3\xi q_{ij} + 2\lambda \Lambda_{ij})\chi^{\rm s}_{ij} + 2\mathcal {P}\chi^{\rm VV}_{ij} 
	\end{equation} 
	by using $\mathcal {P} = 2\mu_{\rm B}\gamma_{\rm n}\hbar \braket {r^{-3}}$ ($\mu_{\rm B}$: the Bohr magneton, $\gamma_{\rm n}$: the nuclear gyromagnetic ratio), a dimensionless parameter $\kappa$, a spin-orbit coupling parameter $\lambda$, $\Lambda_{ij} = \sum_{n\neq 0}\frac{\braket {0|L_i|n}\braket {n|L_j|0}}{E_n-E_0}$ (the ground state energy: $E_0$, the excited one: $E_n$), the spin susceptibility $\chi^{\rm s}_{ij}$, and the Van-Vleck susceptibility $\chi^{\rm VV}_{ij} = 2N\mu^2_{\rm B}\Lambda_{ij}$. 
	The term proportional to $\chi^{\rm s}_{ij}$ in Eq. (3) consists of isotropic core polarization, $3d$ dipole interactions, and a cross term with spin-orbit coupling. 

	To determine $K_{ij}$ and $\delta \nu_{ij}$ in the spin-singlet state, $^{51}$V NMR spectra were measured by rotating the single crystal around the $a$, $b^*$, and $b$ axes [defined in Fig. 1(d)] at 300 K, as shown in Fig. 2. 
	A single line at $B_0 || c$ indicates that all V sites are equivalent, simultaneously satisfying $\delta \nu = \nu_{ZZ}(3\cos^2\theta_0 -1)=0$ ($\theta_0 = 54.7^\circ$: the magic angle). 
	The spectra split into several lines at the other angles due to appearance of inequivalent V sites with clear quadrupole splittings: these are assigned to two or three sets of seven lines, as traced with fitting curves. 
	It indicates symmetry lowering from the original $R\bar{3}m$ structure where all V sites are equivalent at arbitrary $B_0$ directions. 
	In the $R\bar{3}m$ structure with the local trigonal distortion, the principal $Z$ axis having a maximum $\delta \nu$, should be parallel to the $c$ axis, as known in V$_2$O$_3$.\cite{Takigawa} 
	At 300 K, however, we observe a $\delta \nu$ maximum around $\theta_0$ rotated from $c$ to $b^*$ axis and $\pm \theta_0$ from $a$ to $c$ axis, where $\theta_0$ matches the V-O bond direction in the VO$_6$ octahedron. 
	In addition, identical spectral profiles between the $cb^*$ and $cb^\prime$ rotations ($b^\prime$ is located at 120$^\circ$ inclined from $b^*$) indicate that the observed three V sites are connected by three-fold rotational symmetry expected from the in-plane superlattice ${\surd 3}a \times {\surd 3}a$ of the x-ray and electron diffractions.\cite{Cardoso, Takei, Imai, Tian}  

	\begin{figure}
	\includegraphics[scale=0.53]{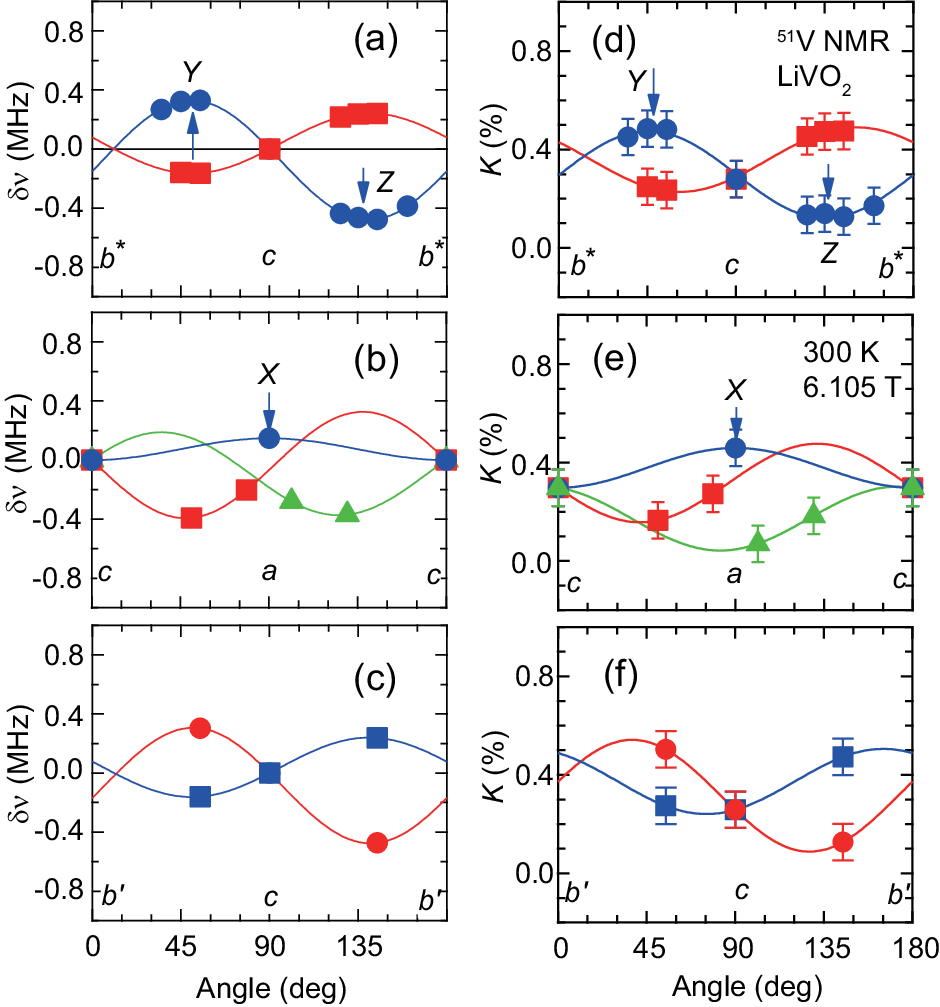}
	\caption{\label{Fig3} 
	(Color online)  Angle dependence of (a,b,c) nuclear quadrupole splitting frequency $\delta \nu$ and (d,e,f) Knight shift $K$ obtained for the $b^*c$, $ca$, and $b^\prime c$ rotations. 
	$X$ and $Z$ are defined by a maximum and minimum of $K$ and $\delta \nu$, respectively. 
	Experimental data are well fitted into the Volkov's sinusoidal formula, $y = y_0^\alpha + y_1^\alpha\cos2(\theta-\phi^\alpha)$ ($y= \delta \nu$, $K$), with fitting parameters $y_0^\alpha$, $y_1^\alpha$ and $\phi$ for the rotation axes $\alpha = a, b^*$, and $b$.\cite{Volkov} 
	}
	\end{figure}

	We define $\delta \nu$ as an average of the splitting frequency independently fitted to the sinusoidal function and plot as a function of the rotation angle in Fig. 3. 
	$\delta \nu$ obtained for the three V sites are well fitted to the Volkov's formula\cite{Volkov} and give the principal components, $(|\delta\nu_{XX}|, |\delta\nu_{YY}|, |\delta\nu_{ZZ}|) = (0.15, 0.33, 0.48)$MHz, where $X$, $Y$, and $Z$ are parallel to $x-y$, $x+y$, and $z$ in Fig. 1(d) and satisfy the three-fold rotational symmetry. 
	The absolute values are good agreement with those for the powder sample, (0.15, 0.33, 0.47)MHz.\cite{Kikuchi} 
	The symmetry $Z$ axis parallel to the VO bond $z$ direction in Fig .1(d) indicates that the predominant $d$ (electron or hole) orbital lies down perpendicular to the $Z$ axis, i. e., has a $d_{X^2-Y^2}$ ($d_{xy}$) character. 
	In the $d^2$ case, a minority orbital (a hole orbital) can contribute to the EFG. 
	Namely, electrons are mainly occupied in $d_{YZ}$ ($d_{yz}$) and $d_{ZX}$ ($d_{zx}$), and the sign of $\delta\nu_{ii}$ should be selected as $(\delta\nu_{XX}, \delta\nu_{YY}, \delta\nu_{ZZ}) = (-0.15, -0.33, 0.48)$MHz. 
	In a point-charge approximation, $\delta \nu^{\rm out}$ is obtained as $(\delta\nu_{XX}, \delta\nu_{YY}, \delta\nu_{ZZ}) = (-0.01, 0.04, 0.05)$MHz for $\gamma_\infty$, based on the crystal structure,\cite{Pourpoint} where the $Z$ axis is close to the $c$ axis due to the residual trigonal distortion. 
	The deviation of the $Z$ axis from the experimental results suggests the negligible $\delta \nu^{\rm out}$ contribution likely due to the small $1-\gamma_\infty$ in the present case. 
	From Eq. (1), the orbital occupation is evaluated as $d_{X^2-Y^2} (d_{xy}) : d_{YZ} (d_{yz}) : d_{ZX} (d_{zx}) = 0.34 : 0.77 : 0.89$. 
	The result does not contradict to the $d_{yz}d_{zx}$ orbital order accompanied with the trimer-type valence bond order proposed theoretically by Pen {\it et al.}\cite{Pen} but shows a significant contribution of $d_{xy}$. 

	Since $\chi^{\rm s}$ already goes to zero at 300 K, the anisotropy in $K$ is governed by the Van-Vleck term of Eq. (3). 
	The symmetry of the $3d$ orbital state is reflected in the anisotropic Van-Vleck term via the second-order mixing of the ground and excited states under the external magnetic field. 
	$K$ defined as a relative shift of the central line is also plotted in Fig. 3. 
	The angular dependences of $K$ exhibit similar profiles to those of $\delta \nu$ for the three rotations. 
	The principal components $(K_{XX}, K_{YY}, K_{ZZ}) = (0.46, 0.47, 0.12)\pm0.02 \%$ give an isotropic part $K_{\rm iso} = (K_{XX} + K_{YY} + K_{ZZ})/3$ = 0.35\% corresponding to $\chi^{\rm VV}_{\rm iso} = 4.9 \times 10^{-5}$ emu/V-mol, using the orbital hyperfine coupling $-2\mathcal {P}  = 40$ T (Ref. \onlinecite{Abragam}) for V$^{3+}$. 
	The $Z$ axis is identical to that of $\delta \nu$, consistent with the predominant $d_{xy}$ hole state [Fig. 1(c)]. 

	\begin{figure}
	\includegraphics[scale=0.7]{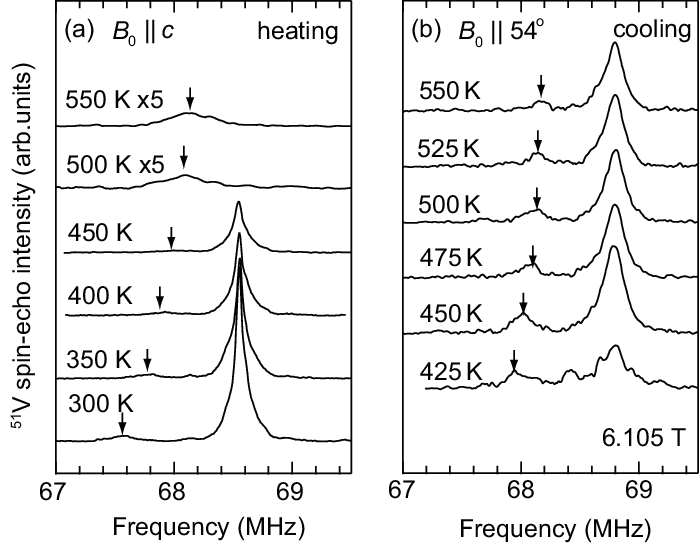}
	\caption{\label{Fig5} 
	(Color online) Temperature dependence of the $^{51}$V NMR spectra measured at (a) the $c$ axis and (b) 54$^\circ$ from the $c$ axis in LiVO$_2$. 
	The spectrum pointed by the arrow comes from the impurity LiV$_2$O$_4$. 
	}
	\end{figure}

	To gain further insights into the phase transition, $^{51}$V NMR spectra were measured across $T_c$ at the $c$ axis and $\theta_0$, as shown in Fig. 4. 
	A low-frequency (67.5 MHz at 300 K) peak with a negative shift comes from an impurity phase of LiV$_2$O$_4$.\cite{Shimizu4} 
	For $B_0 || c$, a sharp peak is observed at 68.5 MHz and suddenly disappears due to fastening of the spin-echo decay time $T_2$ on heating above $T_c \sim$ 500 K. 
	Such abrupt fastening of $T_2$ except for the magic angle is also encountered in correlated paramagnetic metals V$_2$O$_3$ and LiV$_2$O$_4$.\cite{Takigawa, Shimizu4} 
	Alternatively, multiple peaks observed for $B_0 || \theta_0$ become single above 450 K. 
	A difference of $T_c$ in the two measurements arises from thermal hysteresis of the first-order transition. 
	Such a switching of the magic angle indicates a change in the symmetry axis of the $d$ orbital across $T_c$. 
	The magic angle of $\theta_0$ above $T_c$ agrees with the trigonal symmetry,\cite{Takigawa, Shimizu4} as expected from the crystal structure. 

	From the resonance line at the magic angles in Fig. 4, we obtained the Knight shift of the paramagnetic state as $K_{\rm iso}$ = 0.67\%, which is larger than that of the nonmagnetic state, 0.35\%. 
	Here $K_{\rm iso}$ is expressed as $\mathcal {P} (\kappa + 2\lambda \Lambda_{ii})\chi^{\rm s}_{ii} + 2\mathcal {P}\chi^{\rm VV}_{ii}$ from Eq. (3). 
	Since $\kappa + 2\lambda \Lambda_{ii}$ would be negative for vanadates owing to the predominant core polarization, an increase in $K$ above $T_c$ indicates the appreciable spin-orbit coupling term $2\lambda \Lambda_{ii}\chi^{\rm s}_{ii}$ and the Van-Vleck term $2\mathcal {P}\chi^{\rm VV}_{ii}$. 
	The hyperfine coupling constant ($-6$ T/$\mu_{\rm B}$)\cite{Kikuchi} is indeed much larger than that of V$_2$O$_3$ ($-13$ T/$\mu_{\rm B}$)\cite{Takigawa}. 
	The enhanced $\chi^{\rm VV}_{ii}$ comes from the residual degeneracy in the three-fold $t_{2g}$ levels, as expected from the calculated orbital occupations in the paramagnetic state.\cite{Ezhov} 
	Using $\mathcal {P} (\kappa + 2\lambda \Lambda_{ii}) = -6$ T/$\mu_{\rm B}$ (Ref. \onlinecite{Onoda}) and $-2\mathcal {P}  = 40$ T/$\mu_{\rm B}$ (Ref. \onlinecite{Abragam}), we obtained $K^{\rm VV} = 2.7\%$ and $\chi^{\rm VV} = 3.6 \times 10^{-4}$ emu/mol, which corresponds to 47\% of the bulk magnetic susceptibility ($7.6 \times 10^{-4}$ emu/mol).\cite{Tian} 
	By subtracting $\chi^{\rm VV}$ from the bulk susceptibility, we obtained $\chi^{\rm s} \sim 4.0 \times 10^{-4}$ emu/mol. 
	This value is even lower than that of the metallic phase V$_2$O$_3$ (Ref. \onlinecite{Takigawa}) and suggests the itinerant nature of the paramagnetic phase of LiVO$_2$.  

	The above results have revealed the local orbital structures in the triangular lattice LiVO$_2$ across the structural transition by the investigation of the electrostatic and magnetic hyperfine couplings. 
	In the paramagnetic state, the orbital state is governed by the trigonal crystal field that remains degeneracy in $e_g^\prime$. 
	The moderate value of the spin susceptibility suggests the system close to a metal-insulator crossover or a metallic region.\cite{Katayama} 
	Below the first-order structure transition temperature, we observed the $d_{xy}$ hole orbital order that makes occupied $d_{yz}d_{zx}$ spins inert due to the strong singlet formation via the direct $d$-$d$ bonding. 
	This trend is promoted by the large displacement of V atoms. 
	The V-V distance (2.56\AA ) is shorter than that of V metals (2.61\AA ) and VO$_2$ (2.65\AA ),\cite{Schmidt} suggesting the metallic bonding formation. 
	Thus, the present orbital oder appears in order to gain the intersite $d$-$d$ exchange energy or form the trimer orbital, instead of the Jahn-Teller distortion. 
	Unlike the $d^1$ spin-dimer insulators such as MgTi$_2$O$_4$ and VO$_2$, the orbital polarization in $d^2$ system indicates that each vanadium site has a spin $S \sim 1$, similar to the Mott insulator.\cite{Pen} 
	However, the spin gap (1500-3400 K)\cite{Tanaka, Kikuchi} in LiVO$_2$ is too large for the spin-Peierls Mott state where the spin gap should be much smaller than the charge gap, but comparable to the charge gap.\cite{Tian} 
	Taking these experimental data into account, one can regard the system as an orbitally polarized band insulator. 
	Namely, the structural phase transition is the orbital-assisted Peierls transition.\cite{Schmidt} 
	The appearance of the correlated band insulator agrees with the numerical calculations of the Hubbard model.\cite{Yoshitake, Kancharla} 
	Although the orbital occupation is much larger than the theoretical prediction,\cite{Yoshitake} the V displacements that was not taking into account in the calculation may enhance orbital polarization. 

	In the band insulator without orbital degrees of freedom, one expects that a band gap or a bonding-antibonding orbital band transition is equivalent to a spin gap. 
	For the band insulator with the gap of thousands Kelvin, the orbital susceptibility should be vanishingly small ($\chi^{\rm VV}_{\rm iso} < 10^{-6}$ emu/V-mol). 
	The observed residual ($\chi^{\rm VV}_{\rm iso} = 4.9 \times 10^{-5}$ emu/V-mol) suggests the low-lying inter-orbital excitation on the single ion. 
	The different energy structure in inter-orbital correlations may be characteristic in the orbitally polarized Peierls state and provide important insights into the Mott versus Peierls arguments in transition metal compounds such as VO$_2$.\cite{Wentzcovitch} 

	In conclusion, we addressed the microscopic origin for the anomalous spin-singlet formation in LiVO$_2$ with the triangular lattice. 
	Our orbital-resolved NMR study revealed the $d$ orbital occupation across the structural phase transition. 
	We observed the $d_{xy}$ hole order with the three-fold symmetry in the Knight shift and nuclear quadrupole coupling for the spin-singlet state. 
	The rotation of the local symmetry axis gives microscopic evidence for the orbital reformation across $T_c$. 
	The origin of the large spin gap arises from the direct $d$-$d$ $\sigma$ bonding between $d_{yz}$ and $d_{zx}$ orbitals in the vanadium trimer. 
	In this respect, the phase transition is driven by the strong electron-phonon coupling and orbital ordering to relieve the geometrical frustration and effectively gain the intersite exchange energy, which can be regarded as the two-dimensional Peierls transition. 
	The residual low-energy orbital excitation may feature the orbitally-polarized band insulator with moderate electron correlations. 

	We thank technical assistance by S. Inoue and discussion with Y. Motome and T. Yoshitake. 
	This work was financially supported by the Grants-in-Aid for Scientific Research (No.22684018, 23225005, and 24340080) from JSPS, and the Grant-in-Aid for Scientific Research on Priority Areas "Novel State of Matter Induced by Frustration" (No. 22014006) from the MEXT.

\end{document}